\documentclass[ twocolumn,
aps,prd,   
               preprintnumbers,numbers,sort&compress,nofootinbib,
                            showpacs,
               colorlinks,
               linkcolor=blue,   
               citecolor=blue]{revtex4-1}
   \newcommand{\exclude}[1]{}

\usepackage{graphicx,amsmath,amssymb,bm}
\usepackage{psfrag}
\usepackage{feynmp}
\usepackage{hyperref}
\usepackage{enumitem}

\newcommand{\beq}{\begin{equation}}
\newcommand{\eeq}{\end{equation}}
\newcommand{\be}{\begin{eqnarray}}
\newcommand{\ee}{\end{eqnarray}}

\def\dd{ \,\mathrm{d} }

\def\+{\dagger}
\def\la{\langle}
\def\ra{\rangle}
\def\<{\langle}
\def\>{\rangle}

\newcommand{\Lbar}{\Lambda_{\overline{\mathrm{QCD}}}}
\newcommand{\qcd}{{\overline{\mathrm{QCD}}}}

\newcommand{\Tr}{\mathrm{Tr}}

\begin{document}


\title { Dynamical de Sitter phase and nontrivial holonomy in strongly coupled gauge theories in expanding Universe.}

\author{   Ariel R. Zhitnitsky} 
 \affiliation{Department of Physics \& Astronomy, University of British Columbia, Vancouver, B.C. V6T 1Z1, Canada}


\begin{abstract}
 We discuss a new scenario for early cosmology when  the inflationary de Sitter phase emerges dynamically. 
This genuine quantum effect occurs as a result of   dynamics of the topologically nontrivial sectors in a   strongly coupled QCD- like gauge theory   in an expanding universe. 
We test these ideas by explicit computations in hyperbolic space $ \mathbb{H}^3_{\kappa}\times \mathbb{S}^1_{\kappa^{-1}}$.
We argue that the key element for this idea to work is the presence of nontrivial holonomy computed along  $\mathbb{S}^1_{\kappa^{-1}}$. The effect is non-local in nature,  non-analytical in coupling constant and can not be described in terms of any local propagating degree of freedom such as  scalar inflaton field $\Phi(x)$. We discuss some profound phenomenological consequences of this scenario for  inflationary cosmology.  We also suggest to test  these ideas   in a tabletop experiment by measuring some specific corrections to the  Casimir pressure in  the Maxwell theory  formulated  on a topologically nontrivial manifold.   
 \end{abstract}

\maketitle

\section{Introduction. Motivation.}
The main motivation for the present studies is the proposal that inflationary de Sitter phase
\cite{inflation, linde,mukhanov}
 may be  dynamically generated 
as a  result of   presence of the  topologically  nontrivial  sectors in expanding universe. Inflaton in this framework \cite{Zhitnitsky:2013pna,Zhitnitsky:2014aja} is an auxiliary topological non-propagating field with no canonical kinetic term, similar to known topologically ordered phases in condensed matter systems. This auxiliary field effectively describes the dynamics of the topological sectors $|k\ra$ in a gauge theory (coined as $\qcd$ in \cite{Zhitnitsky:2013pna,Zhitnitsky:2014aja})  in  expanding Universe.  

This picture  should be contrasted with conventional proposals reviewed in \cite{linde,mukhanov} when the de Sitter behaviour is achieved in quantum field theory (QFT) by  
  assuming the existence of a new scalar local field  $\Phi (x)$ with a non-vanishing potential energy density $V(\Phi)$. The shape of this potential energy can be adjusted in a such a way that the  contribution to energy density $\epsilon$ and pressure  $p$  is in agreement with observations. In different words, the scale parameter ${\rm{a}} (t)$ and 
the equation of state during the inflation take the following  approximate  form, 
 \be
\label{a}
{\rm{a}}(t)\sim \exp (Ht), ~~~ \epsilon\approx  -p.
\ee

The key ingredient of the proposal \cite{Zhitnitsky:2013pna,Zhitnitsky:2014aja} is a conjecture that the   vacuum energy in context of the Friedmann-Lema\^itre-Robertson-Walker (FLRW) Universe has the following expansion at small $H\ll \Lbar$
\be	\label{FLRW}
  E_{\mathrm{FLRW}}(H)\sim \left[\Lbar^4+ H\Lbar^3+ {\cal{O}}(H^2)\right],
\ee
when the first non-vanishing term is linear $\sim H$, rather than (commonly accepted) quadratic $\sim H^{2}$  in the Hubble constant. If this conjecture turns out to be correct, than  the Friedman equation assumes the   form
\be	
\label{friedman-infl}
  H^2 &\simeq& \frac{8\pi G}{3} \Delta E,  ~~\Longrightarrow ~~~ H_0\simeq \frac{8\pi G}{3} \Lbar^3\\
   \Delta E&\equiv&   \left[E_{\mathrm{FLRW}}(H)-E_{\mathrm{Mink}}\right]\sim H.\nonumber
\ee
which automatically leads to a non-trivial solution with constant $H_0$, and as a consequence, to a desired de Sitter  behaviour  (\ref{a}).  

There are two  critical elements   in writing equation (\ref{friedman-infl}). First one, as we already mentioned,   is related to the expansion (\ref{FLRW}),  see few  comments on this conjecture below. The second key element 
is   a paradigm  that the relevant definition of the energy in an expanding background which enters the Friedman   equation is the difference $\Delta E (H)\equiv \left[E(H)-E_{\mathrm{\rm Mink}}\right]$, similar to the computation   of the Casimir pressure when the observable energy is the difference similar to $ \Delta E$. 
This element in our analysis is not a new proposal  identifying  $ \Delta E$ with gravitating energy from  the Friedman   equation.  In fact, in the present context such a definition for the vacuum energy was advocated long ago in 1967 by Zeldovich \cite{Zeldovich:1967gd} for the first time. Later on such  definition for the relevant energy $\Delta E\equiv (E_{\rm FLRW} -E_{\rm Mink})$ which  enters the Friedman  equations has been advocated from   different perspectives in a number of papers,  see e.g.  relatively recent works~\cite{Bjorken:2001pe, Schutzhold:2002pr, Klinkhamer:2007pe, Thomas:2009uh,Maggiore:2010wr}, see also review article \cite{Sola:2013gha} with 
  large number of references on original papers. 
 Essentially, this prescription implies that $ \Delta E$ may only depend on properties of the external gravitational background, while the conventional contributions computed in Minkowski  flat space-time (such as the QCD vacuum energy or the Higgs potential in electroweak theory)   are automatically subtracted by this prescription\footnote{A  somewhat similar, but not identically the same  subtraction procedure  has been suggested recently in refs. \cite{Kaloper:2013zca,Kaloper:2014dqa}, the so-called ``vacuum energy sequestering" proposal.  The prescription \cite{Kaloper:2013zca,Kaloper:2014dqa}  is also inherently non-local, similar to the crucial role of non-locality in our framework realized  in terms of the holonomy (\ref{holonomy}). In fact, our computation of the vacuum energy (\ref{FLRW}) as discussed below, is based on evaluation of the holonomy (\ref{holonomy}) along the entire history of the universe, which resembles in spirit   the computations of the so-called ``historic averages" in refs.  \cite{Kaloper:2013zca,Kaloper:2014dqa}. Furthermore, we have to keep the volume of the system to be finite in the computations for the infrared regularization of the theory. It is akin to that from refs.\cite{Kaloper:2013zca,Kaloper:2014dqa} where the finite volume is also  required property for the consistency of the procedure.}. We shall not elaborate on a number of subtle points related to this prescription in the present work referring to the original papers and review article 
 \cite{Sola:2013gha}.

The main topic  of the preset paper  is analysis of another  critical element, briefly mentioned above, and  leading to   (\ref{friedman-infl}). There is  well known,  conventional  and  generally accepted argument which suggests  that the expansion (\ref{FLRW}) starts with a quadratic   $\sim H^{2}$, rather than the liner $\sim H$ term. The argument   is based on fundamental principles of    {\it locality} and general covariance, see original papers \cite{Shapiro:1999zt,Shapiro:2000dz},  recent review \cite{Sola:2013gha}, and some comments \cite{Zhitnitsky:2013pna} with pros and cons of these arguments. Indeed, the curvature scalar $R$ for FLRW Universe
is quadratic in H,
\be
\label{R}
|R|=6 \left(\frac{\ddot{\rm{a}}}{{\rm{a}}}+\frac{\dot{\rm{a}}^2}{{\rm{a}}^2}\right)=12H^2+6\dot{H}, 
\ee
when $\dot{H}\sim {\cal{O}}(H^2)$, see  \cite{Sola:2013gha}. 
Therefore, if the infrared (IR) behaviour of the system is entirely determined   by the local characteristics, such as curvature scalar $R$ and/or  higher order derivative terms $R^2, R_{\mu\nu}R^{\mu\nu}$, than the corrections to the energy (\ref{FLRW})
indeed must be proportional to  even powers $H^{2n}$ as  correctly argued in  \cite{Shapiro:1999zt,Shapiro:2000dz,Sola:2013gha}. 

However, the main essence of the proposal  \cite{Zhitnitsky:2013pna,Zhitnitsky:2014aja} is precisely the observation  that the conventional 
assumption on locality might be badly violated in strongly coupled gauge theories.  
The basic reason for such violation is well known
and well-understood,   at least in Minkowski space-time. The energy (\ref{FLRW})  is generated  due to the  tunnelling events between $|k\ra$  topological sectors, which formulated in terms of inherently  {\it non-local}   large gauge transformation operator $\cal{T}$.  
Furthermore, this energy  has non-dispersive nature, i.e. it can not be formulated in terms of any {\it local} propagating degrees of freedom\footnote{\label{top}This energy can be expressed  in terms of the contact term in the topological susceptibility, determined by the IR physics and boundary conditions. The corresponding physics    has been well understood  using  the lattice numerical simulations in strong coupling regime, see  \cite{Zhitnitsky:2013pna} for references and details. }.   Transition from 
Minkowski space-time to   time dependent background (\ref{FLRW})  obviously will   not    modify  the   nature and origin of this type of energy. Rather, a transition to FLWR Universe  introduces some   background- dependent corrections  to the same type of    energy (\ref{FLRW}),
which was coined as ``strange energy" in  \cite{Zhitnitsky:2013pna,Zhitnitsky:2014aja} due to its unconventional origin as mentioned above.   

One should comment here that this  feature of {\it non-locality} when the system is not completely characterized by a  local physics is very  similar to the well known property in topologically ordered phases  in condensed matter physics wherein an expectation value of a local operator does not fully characterize the ground state of the system.
Instead, one should use  some {\it non-local} variables for proper characterization of the system.

The main subject of the present work   is to elaborate and clarify a number of non-trivial questions related to the non-locality in QFT  and generation of the linear $\sim H$ term in (\ref{FLRW})  in some simplified models\footnote{\label{H}Here and in what follows we use term ``linear in $H$ correction" as a generic feature to distinguish a nontrivial  background from the trivial Euclidean space. This dimensional parameter should not be literally identified with the Hubble constant. Rather, it could be any other dimensional parameter which characterizes the system, such as the size of torus related to the nontrivial holonomy (\ref{holonomy}) with $\beta\sim H^{-1}$.}.   The basic point of our discussions is that a gauge QFT  (when one should sum over all topological sectors $|k\ra$ in the definition of the partition function)  
is not fully described by the local characteristics, such as curvature (\ref{R}). In particular, the linear dependence  on the  background may enter 
(\ref{FLRW}) through other  characteristics of the system such as holonomy 
\be
\label{holonomy}
U(\mathbf{x})={\cal{P}}\exp\left(i\int_0^{\beta} dx_4 A_4(x_4, \mathbf{x})\right),
\ee
which is gauge invariant but non-local object as it depends on the boundary conditions. We shall argue below  that precisely the  non-trivial holonomy in gauge theories plays  a key role in the mechanism which could generate  the  ``strange energy" (\ref{FLRW}).  It is very hard technical problem to compute the non-perturbative energy (\ref{FLRW}) in a time- dependent background characterized by parameter $H$, see footnote \ref{H} for clarification. However, one can simplify the problem by considering the sensitivity of a gauge system to  some  external  dimensional parameters   characterizing    the  gravitational  background, such as $\kappa$, see definition below. This  parameter   plays   a  role similar  to the Hubble constant $H$ in  FLRW  Universe (\ref{FLRW}).  Our goal is to study the dependence of  the  ``strange energy" (\ref{FLRW}) as a function of $\kappa$  in the limit of small $\kappa\rightarrow 0$   in some simple settings   where such computations can be performed.

The basic idea is as follows. We would like to consider hyperbolic space $ \mathbb{H}^3_{\kappa}$  with the constant negative curvature 
$-\kappa^2 $. As  we discuss below, there is a conformal equivalence between $(\mathbb{R}^4- \mathbb{R}^2) $ and  $ \mathbb{H}^3_{\kappa}\times \mathbb{S}^1_{\kappa^{-1}}$ where  $\mathbb{S}^1_{\kappa^{-1}}$ denotes the circle of radius $\kappa^{-1}$. The holonomy (\ref{holonomy}) is computed precisely along a closed loop $\mathbb{S}^1_{\kappa^{-1}}$. Our goal is to study the first nontrivial correction $\sim \kappa$ to the nonperturbative energy  (\ref{FLRW}) in the limit of small $\kappa\rightarrow 0$ corresponding to smooth transition to   Euclidean space $\mathbb{R}^4$. We should recover the Euclidean results    when $\kappa$ identically  vanishes. 

The key observation is that  the topological configurations with non-trivial holonomy 
(\ref{holonomy})  produce a finite contribution to the energy density (\ref{FLRW}) with corrections being  linearly proportional to $\kappa$. Such effects can not be expressed in  terms of any local operators such as curvature (\ref{R}). Rather, it is generated due to topological  vacuum configurations with nontrivial holonomy (\ref{holonomy}), not expressible in terms of local observables. This is precisely the reason why the generic arguments \cite{Shapiro:1999zt,Shapiro:2000dz,Sola:2013gha}
 based on locality simply do not apply here.
 
 Therefore, our computations   of the linear correction $\sim \kappa$ in the vacuum energy density  using simplified model with $\mathbb{H}^3_{\kappa}\times \mathbb{S}^1_{\kappa^{-1}}$ background can be thought  as a strong argument  
 supporting our conjecture on the liner correction $\sim H$ in generic FLRW  Universe (\ref{FLRW})-- in both cases the linear correction is not associated with the local curvature operator (\ref{R}). 
   
   One may wonder how a   bulk property (such as vacuum energy density)  in a  gapped theory  could be ever sensitive to such global characteristics as radius of the circle of $\mathbb{S}^1_{\kappa^{-1}}$? The answer lies not in the local, but global properties of the space. 
   Imagine that we study the Aharonov-Casher effect.
We insert an external charge into a superconductor in which the electric field is exponentially suppressed $\sim \exp(-r/\lambda)$ with $\lambda $ being the penetration depth.
Nevertheless, a neutral magnetic fluxon will be still sensitive to an inserted external charge at arbitrary large distances in spite of the screening of the physical field (which is equivalent to the presence of a gap in our system).
This genuine quantum effect is purely topological and non-local in nature and can be explained in terms of the dynamics of the gauge sectors which are responsible for the long range dynamics.
Imagine now that we study the same effect but in a different background. 
The corresponding topological sectors will be modified due to the variation of the external background.
However, this modification can not be described in terms of any local dynamical fields, as there are no any propagating long range fields in the system since the physical electric field is screened.
For this simplified example, the dynamics of the ``strange energy" as a function of $\kappa$ is determined by  the modifications of topological sectors when the background varies.  
The effect is obviously non-local in nature as the Aharonov-Casher effect itself is a non-local phenomenon.

 \exclude{Our analysis presented below  in  a simplified setting when the system is  formulated on $ \mathbb{H}^3_{\kappa}\times \mathbb{S}^1_{\kappa^{-1}}$ is in fact not so far away from reality with (almost) de Sitter universe, which can be considered as a constant (positive) curvature surface embedded in 5d Minkoski space time. At the same time the  $\mathbb{H}^3_{\kappa}$ space can be considered as 
a  constant (negative)  curvature surface embedded in 4d Euclidean  space. The transition between positive and negative curvatures can be always achieved using analytical continuation \cite{mukhanov}.  As we discuss in concluding section \ref{conclusion} the presence of nontrivial $ \mathbb{S}^1$ which is a required feature of the space-time  to produce a nontrivial holonomy (\ref{holonomy})    is consistent with all known observations if the size of the  system 
 is finite but sufficiently large $ \gtrsim H^{-1}$. }
 
 The readers  interested   in the cosmological consequences, rather than in technical computational  details may directly jump  to section \ref{energy} where we list the main results of this framework.  Section \ref{conclusion} is our conclusion where we suggest to test some of the ideas presented in this work   in a tabletop experiment by measuring some specific corrections to the Casimir vacuum energy in the Maxwell theory formulated on 4-torus with the nontrivial holonomy.   
 
\section{\label{review}The nature  of   ``strange"  energy (\ref{FLRW}) in Euclidean space $\mathbb{R}^4$}  
The main goal of this  section is to review a number of  crucial  elements relevant for our studies. 
We start in section \ref{contact} with   explanation of  a highly nontrivial nature of the ``strange energy" (\ref{FLRW})  in the Euclidean space time.   We continue in section \ref{sec:holonomy},  by clarifying  the crucial role of the holonomy (\ref{holonomy}) in generating such type of energy. We present few technical results in sections \ref{sec:details}, \ref{ensemble}. Finally, in section \ref{interpretation} we elaborate  on non-local features of relevant vacuum configurations saturating the ``strange energy"   in cosmological context. The corresponding analysis will play an important role in our main   section \ref{sec:hyperbolic} when we 
study the hyperbolic space $\mathbb{H}^3_{\kappa}\times \mathbb{S}^1_{\kappa^{-1}}$ and analyze the properties of the ``strange" energy as a function  of $\kappa$ at small $\kappa\rightarrow 0$.

\subsection{\label{contact}The topological susceptibility and contact term} 
We start our  short   overview on  the ``strange" nature of the vacuum energy (\ref{FLRW}) by reviewing a  naively unrelated topic--  the formulation and resolution of the so-called $U(1)_A$ problem in strongly coupled QCD~\cite{witten,ven,vendiv}. We 
   introduce  the topological susceptibility $\chi$ which is ultimately related to the vacuum energy $E_{\mathrm{vac}}(\theta=0)$ 
   as follows\footnote{We use the Euclidean notations  where  path integral computations are normally performed.}
\be
\label{chi}
 \chi =    \left. \frac{\partial^2E_{\mathrm{vac}}(\theta)}{\partial \theta^2} \right|_{\theta=0}= \lim_{k\rightarrow 0} \int \!\dd^4x e^{ikx} \la T\{q(x), q(0)\}\ra  ~~ 
 \ee
where     $\theta$ parameter   enters the  Lagrangian   along with  topological density operator $q (x)=
\frac{1}{16 \pi^{2}} \mathrm{tr}[ F_{\mu\nu} \tilde{F}^{\mu\nu}]$ and $E_{\mathrm{vac}}(\theta)$ is the ``strange" vacuum energy density
which  represents  the first   term in  expansion (\ref{FLRW}) corresponding to the flat space-time background. This $\theta$- dependent portion of the vacuum energy (computed  at $ \theta=0$)   has a number of unusual properties as we review below.  
The corresponding properties  are  easier  to explain in terms of the correlation function (\ref{chi}), rather than in terms of the vacuum energy $E_{\mathrm{vac}}(\theta=0)$ itself.  The relation between the two  is given by eq. (\ref{chi}).
 
First of all,  the topological susceptibility $\chi$  does not vanish in spite of the fact that $q= \partial_{\mu}K^{\mu}$ is total divergence. This feature is very different from any conventional correlation functions  which normally must  vanish  at zero momentum if the  corresponding operator  can be represented as 
total divergence.  

Secondly, any physical $|n\ra$   state gives a negative contribution to this 
diagonal correlation function
\be	\label{G}
  \chi_{\rm dispersive} \sim  \lim_{k\rightarrow 0} \int d^4x e^{ikx} \la T\{q(x), q(0)\}\ra \nonumber \\
  \sim 
    \lim_{k\rightarrow 0}  \sum_n \frac{\la  0 |q|n\ra \la n| q| 0\ra }{-k^2-m_n^2}\simeq -\sum_n\frac{|c_n|^2}{m_n^2} \leq 0,  
\ee
 where   $m_n$ is the mass of a physical $|n\ra$ state,  $k\rightarrow 0$  is  its momentum, and $\la 0| q| n\ra= c_n$ is its coupling to topological density operator $q (x)$.
 At the same time the resolution of the $U(1)_A$ problem requires a positive sign for the topological susceptibility (\ref{chi}), see the original reference~\cite{vendiv} for a thorough discussion, 
\be	\label{top1}
  \chi_{\rm non-dispersive}= \lim_{k\rightarrow 0} \int \!\dd^4x e^{ikx} \la T\{q(x), q(0)\}\ra > 0.~~~
\ee
Therefore, there must be a contact contribution to $\chi$, which is not related to any propagating  physical degrees of freedom,  and it must have the ``wrong" sign. The ``wrong" sign in this paper implies a sign 
  which is opposite to any contributions related to the  physical propagating degrees of freedom (\ref{G}). 
  The ``strange energy" in this paper implies the $\theta$ dependent portion of the energy (\ref{FLRW}), (\ref{chi}) which {\it can not} be formulated  in terms of conventional propagating degrees of freedom as it has pure non-dispersive nature according to eqs. (\ref{G}), (\ref{top1}).

    In the framework \cite{witten} the contact term with ``wrong" sign  has been simply postulated, while in refs.\cite{ven,vendiv} the Veneziano ghost (with a ``wrong" kinetic term) had been introduced into the theory to saturate the required property (\ref{top1}).   
    
  Third,  the  contact term (\ref{top1}) has  the structure $\chi \sim \int d^4x \delta^4 (x)$.
  The significance of this structure is  that the gauge variant correlation function in momentum space
  \be
  \label{K}
   \lim_{k\rightarrow 0} \int d^4x e^{ikx} \la K_{\mu}(x) , K_{\nu}(0)\ra\sim   \frac{k_{\mu}k_{\nu}}{k^4}
   \ee 
  develops  a topologically protected  ``unphysical" pole which does not correspond to any propagating massless degrees of freedom, but nevertheless must be present in the system. Furthermore, the residue of this   pole has the ``wrong sign".
  This ``wrong sign" is  due to   the Veneziano ghost  contribution   saturating the non-dispersive term  in gauge invariant correlation function (\ref{top1}),
   \be
  \label{K1}
   \< q({x}) q({0}) \> \sim  \la \partial_{\mu}K^{\mu}(x) , \partial_{\nu}K^{\nu}(0)\ra \sim \delta^4(x).
   \ee 
 We conclude this review-type   subsection with the following remark. The entire framework, including the singular behaviour of
  $ \< q({x}) q({0}) \>$   with the ``wrong sign",  has been well confirmed by numerous  lattice simulations in strong coupling regime, and it is accepted by the community as a standard resolution of the $U(1)_A$ problem. Furthermore, it has been argued long ago in ref.\cite{Luscher:1978rn}
  that the gauge theories may exhibit the ``secret long range forces" expressed in terms of the correlation function (\ref{K}) with topologically protected pole at $k=0$. 
  
  Finally, in a weakly coupled gauge theory (the so-called ``deformed QCD" model \cite{Yaffe:2008}) where all computations  can be performed in  theoretically  controllable way  
one can explicitly test every single element of this entire framework, including the topologically protected pole (\ref{K}), the contact term with ``wrong sign", etc,  see ref. \cite{Thomas:2011ee,Zhitnitsky:2013hs} for the details. In particular, one can explicitly see that the Veneziano ghost is in fact an auxiliary topological field which saturates the vacuum energy and the topological susceptibility $\chi$.  What is more important for the present studies is that  one can explicitly see that the holonomy (\ref{holonomy}) plays a crucial role   in generating the ``strange" vacuum energy and contact term in topological susceptibility. 
  
While all these unusual features of the vacuum energy are well-known and well-supported by numerous lattice simulations (see e.g. \cite{Zhitnitsky:2013hs} for a large number of references on original lattice results) the analytical understanding of these properties   in strong coupling regime is still lacking.  In next subsection we review some known results on this matter specifically emphasizing on role of the holonomy (\ref{holonomy}) in the analytical computations. Precisely a nontrivial holonomy  (\ref{holonomy}) may play a crucial  role in generating  the  linear correction $\sim H$ in eq. (\ref{FLRW}) as we argue in section \ref{energy}. This is the key technical element  which pinpoints the source of linear corrections $\sim H$  not    expressible  in terms of any  local operators such as curvature (\ref{R}). 

In conclusion we  should comment  that 
the vacuum energy in electroweak (EW)  sector of the standard model  is not sensitive to the the topological features of the EW  gauge fields of the Standard Model ($W_{\mu}^{\pm}, Z_{\mu}, A_{\mu}$) as these topological properties play no role in dynamics of these fields, in huge contrast with QCD. This is  due to the fact that the mass gap of  the non abelian  EW gauge bosons is resulted from the Higgs mechanism, in contrast with QCD where the mass gap and the vacuum energy are  dynamically generated by the topological fluctuations with nontrivial holonomy (\ref{holonomy}).   Therefore, the subtraction procedure, formulated in Introduction, automatically removes all the vacuum energy related to the EW  sector of the standard model without any  corrections $\sim H$.   The  linear corrections are specific to the strongly coupled QCD with its topological features formulated in terms of the non-local operators, the holonomy (\ref{holonomy}) and large gauge transformations operator  $\cal{T}$  as discussed above. 

\subsection{\label{sec:holonomy}The  holonomy (\ref{holonomy}) and generation of the ``strange" energy    in Euclidean space.  }
 The key role in our discussions will play the behaviour of holonomy $U(\mathbf{x})$ at spatial infinity,  the Polyakov line,
  \be
\label{polyakov}
L={\cal{P}}\exp\left(i\int_0^{\beta} dx_4 A_4(x_4, |\mathbf{x}|\rightarrow\infty)\right).
\ee
The operator $\Tr L$ classifies the self-dual  solutions which may contribute to the path integral at finite temperature $T\equiv \beta^{-1}$, including the low temperature limit $T\rightarrow 0$.  
There is a well known generalization of the standard self-dual instantons to non-zero temperature, which corresponds to 
 the description on $\mathbb{R}^3 \times \mathbb{S}^1$ geometry. This is so-called periodic instantons, or calorons\cite{Harrington:1978ve}  studied in details in \cite{Gross:1980br}. These  calorons   have trivial holonomy, which implies that the $\Tr L$   assumes values belonging to the group centre $\mathbb{Z}_N $ for the $SU(N)$ gauge group.  

More general class of the self-dual  solutions with nontrivial holonomy (\ref{polyakov}), the so-called KvBLL calorons  were constructed more  recently  \cite{Kraan:1998pm,Lee:1998bb}. In this case the holonomy (\ref{polyakov}) in general,  is not reduced to the 
group centre $\Tr L\notin \mathbb{Z}_N$. The fascinating feature of the  KvBLL calorons is that they can be viewed as a set of $N $ monopoles of $N$ different types.  Normally, one expects that monopoles come in $N-1$ different varieties carrying a unit magnetic charge from each of the $U(1)$ factors of the $U(1)^{N-1}$ gauge group left unbroken by vacuum expectation value due to nontrivial holonomy (\ref{polyakov}). There is an additional, so called  Kaluza- Klein (KK) monopole which carries magnetic charges and instanton charge. All monopole's charges are such that when complete set of  different types of monopoles are present, the magnetic charges exactly cancel, and the configuration of $N$ different monopoles carries a unit instanton charge.

It has been known since \cite{Gross:1980br}
that the gauge configurations with non-trivial holonomy are strongly suppressed in the partition function. Therefore, naively 
KvBLL calorons can not produce a finite contribution to the partition function. However, this   naive argument is  based on consideration of the individual KvBLL caloron, or finite number of them. If one considers a  grand canonical  assemble of these objects than  their density is determined by the dynamics, and    the old argument  of ref. \cite{Gross:1980br} 
breaks down.  The corresponding objects in this case  may in fact   produce a finite contribution to the partition function.  A self consistent 
computations in a weak coupling regime supporting this picture have been carried out in the so-called ``deformed QCD" model
\cite{Yaffe:2008}. One can explicitly see how $N$ different types of   monopoles with nontrivial holonomy (\ref{polyakov}) which carry fractional topological charge $\pm 1/N$ produce  confinement, generate the ``strange" vacuum  energy (\ref{chi}) and associated with this energy the topological susceptibility (\ref{K1})  with known, but highly unusual properties reviewed above in section \ref{review}, see  \cite{Thomas:2011ee,Zhitnitsky:2013hs} for the technical details on these computations. 

In strong coupling regime we are interested in, the corresponding analytical computations  have never been completed.
There is a limited  number of partial analytical and numerical results  \cite{Diakonov:2004jn,Diakonov:2007nv,Faccioli:2013ja,Shuryak:2013tka,Liu:2015ufa} on computations of moduli space and one loop determinant,  controlling   the dynamics and interaction properties of the  constituents in a large  ensemble of KvBLL calorons. We review these basic technical results in next section \ref{sec:details}, as they will play an important role in our analysis  below.

While complete analytical solution in strong coupling regime is still lacking, nevertheless  there is a number of hints supporting the basic picture  that the KvBLL configurations with nontrivial holonomy (\ref{polyakov}) and representing $N$ different types of monopoles with fractional topological charges $\pm 1/N$ saturate the ``strange" vacuum  energy (\ref{chi}) and associated with this energy the topological susceptibility (\ref{K1}) in a very much the same way as it happens in 
``deformed QCD" model  where all computations are performed in a theoretically controllable regime \cite{Yaffe:2008,Thomas:2011ee,Zhitnitsky:2013hs}. 

In what follows we assume that the ``strange" vacuum  energy (\ref{chi}) and associated with this energy the topological susceptibility (\ref{K1})  is  indeed saturated by  fractionally charged monopoles with $Q=\pm 1/N$ which are constituents of KvBLL caloron with nontrivial holonomy (\ref{polyakov}). With this assumption  in hand the question which is addressed in this work is as follows.  
How does the ``strange energy"   vary in a hyperbolic space 
$\mathbb{H}^3_{\kappa}\times \mathbb{S}^1_{\kappa^{-1}}$ as a function of dimensional parameter $\kappa$? 
The difference between  the original $\mathbb{R}^3 \times \mathbb{S}^1$ and $\mathbb{H}^3_{\kappa}\times \mathbb{S}^1_{\kappa^{-1}}$ spaces is the curvature of the hyperbolic space $R[\mathbb{H}^3_{\kappa}]\sim \kappa^2$ at $\kappa\rightarrow 0$.
If we find a linear dependence on $\kappa$ at small $\kappa$   it would be a strong argument supporting our conjecture (\ref{FLRW}) on linear dependence of ``strange" vacuum energy as a function of external parameter.  Such linear scaling obviously implies that this background-dependent correction is not generated by any local operators such as curvature (\ref{R}), but rather is generated by  nonlocal operator   (\ref{holonomy}), (\ref{polyakov})  which is sensitive to the global characteristics of the background. 

\subsection{\label{sec:details}Nontrivial   holonomy (\ref{polyakov})  in Euclidean space. Few technical details. }
In this section we present few formulae derived   in Euclidean space in order to compare them with parallel  expressions obtained in the hyperbolic space 
$\mathbb{H}^3_{\kappa}\times \mathbb{S}^1_{\kappa^{-1}}$ from  section \ref{sec:hyperbolic}. The corresponding comparison will allow us to study the dependence of the ``strange"  energy as a function of $\kappa$ at small $\kappa\rightarrow 0$.  

We start from analysis of the KvBLL configurations with nontrivial holonomy (\ref{polyakov}). We use $SU(2)$ gauge group in our discussions to simplify notations, though the generalization for $SU(N)$ is also known.  The KvBLL caloron   can be represented as a combination of two monopoles. The first monopole is a conventional BPS monopole, and  at large separations between the constituents in Hedgehog gauge  can be represented as follows
\be
\label{M}
A_4^{M}(r_i)&=&\left(v\coth (v r)-\frac{1}{r}\right) \frac{r_a\tau^a}{2r}\\
A_i^{M} (r_i)&=&\left(1-\frac{vr}{\sinh (vr)}\right)\epsilon_{ijk}\frac{r_j\tau^k}{2r^2}, \nonumber
\ee
 where we adopted the notations from refs \cite{Diakonov:2004jn,Diakonov:2007nv,Faccioli:2013ja,Shuryak:2013tka,Liu:2015ufa,Liu:2015jsa,Larsen:2015vaa} to coin this constituent as $M$-monopole. Parameter $v$ in this formula is arbitrary number  which is determined by  the holonomy (\ref{polyakov}). The classical moduli space is a circle, 
\be
\label{v}
v\in \mathbb{S}^1:~~~ 0\leq v\leq \frac{2\pi}{\beta},
\ee
such that $v$ is an angular variable. For any $v\neq 0$  the gauge group is broken to $U(1)$. In other words, parameter $v$ plays the role of the vacuum expectation value of the Higgs field $\Phi(x)$, which is represented in this system  by $A_4^{M}$-component of the gauge field as expressed by eq. (\ref{M}). One should emphasize that we study gluodynamics, without a scalar Higgs field in the system. Nevertheless, $A_4$- component  of the gauge field  with non vanishing expectation value $v$ plays exactly the same role as the Higgs field in adjoint representation in the standard  BPS equations. Therefore, it is not a mystery that the $r$ dependence of the  $A_4 (r_i)$- component in eq. (\ref{M})  is identically    the same as in conventional BPS construction for the Higgs $\Phi (x) $ field\footnote{\label{normalization}Our  convention for normalization is: $ |\Phi|^2\equiv 2\Tr (\Phi)^2
=2\Tr \left[A_4({r\rightarrow\infty})\right]^2= v^2$}.   

The most important  property of solution (\ref{M}) is the behaviour of $A_4^{M}$ at large distances, which is convenient to represent in the unitary gauge:
\be
\label{M_bc}
A_4^{M}(r\rightarrow\infty)\rightarrow\left(v-\frac{1}{r}\right) \frac{\tau^3}{2}, 
 \ee
 where we also keep the Coulomb like correction $1/r$  along with the leading term $\sim v$ because this Coulomb long range interaction will play an important role in our future discussions.  
One can explicitly see that the holonomy (\ref{polyakov}) is
\be
\label{holonomy1}
\frac{1}{2} \Tr  L =\cos \left(\frac{\beta v}{2}\right) =\cos (\pi\nu), 
\ee
where we introduced the dimensionless parameter $\nu\equiv\frac{\beta v}{2\pi}$. 
The holonomy  belongs to the group  center $\frac{1}{2} \Tr L= \pm 1$ when $v$ assumes its boundary points 
$(v=0,   \frac{2\pi}{\beta})$, in which case   it is called  the  trivial holonomy. 

The second type of monopole is the so-called $L$ monopole which can be constructed from (\ref{M}) as follows \cite{Kraan:1998pm,Lee:1998bb}. First, one should replace $v\rightarrow  \frac{2\pi}{\beta}-v$, which is equivalent to replacement  $\nu\rightarrow (1-\nu)$. Than, one should make a ``large" (improper) gauge transformation 
\be
\label{U}
U_{\rm large} (x_4)=\exp\left(i\frac{\tau_3}{2}\frac{2\pi x_4}{\beta}\right).
\ee
 As it is known the ``large" gauge transformation should be treated differently from ``small" (proper)  gauge transformations 
 because any two field configurations related by   ``large" gauge transformation do not belong to the same gauge orbit. 
 Nevertheless, the transformation (\ref{U}) preserves the periodic boundary conditions because $U_{\rm large} (x_4=0)=
 -U_{\rm large} (x_4=\beta)$. Final step is to perform the reflection $v\rightarrow -v$ in order to restore the original vacuum expectation value (\ref{v}). It is implemented by the discrete transformation $U_{\rm reflection}=\exp(i\tau^2{\pi}/{2})$.  The resulting configuration is the $L$ monopole (the Kaluza-Klein monopole in the original terminology). Its asymptotic behaviour is
 \be
\label{L_bc}
A_4^{L}(r\rightarrow\infty)\rightarrow\left(v+\frac{1}{r}\right) \frac{\tau^3}{2}, 
 \ee
 which should be contrasted with (\ref{M_bc}) with an opposite sign for  a Coulomb term. It   corresponds to the opposite magnetic charges of the  $M$ and $L$ monopoles. Therefore, the action $S$, topological charge $Q$, and magnetic charge $q$ for $M$ and $L$ monopoles are:
 \be
 \label{S}
 S^M&=&\frac{8\pi^2}{g^2}\nu, ~~Q^M= \nu, ~~q^M=+1,~~ \nu\equiv\frac{\beta v}{2\pi},\\
S^L&=&\frac{8\pi^2}{g^2}(1-\nu), ~~ Q^L= (1-\nu), ~~ q^L=-1, \nonumber
 \ee
 while the monopole's mass $m$ is  determined as $S=m\beta$ such that $m=\frac{4\pi}{g^2}v$ when $m$ is expressed in terms of $v$.
 One can explicitly see from (\ref{S}) that the classical action $S=(S^M+S^L)$ for the KvBLL configuration consisting  $L$ and $M$ monopoles does not depend on $v$ and coincides with action  of   the conventional periodic instanton \cite{Harrington:1978ve,Gross:1980br} with $Q^M+Q^L=1$ and action $S={8\pi^2}/{g^2}$. On the quantum level the partition function, of course,  will depend on $v$. We  review the relevant results from refs  \cite{Diakonov:2004jn,Diakonov:2007nv,Faccioli:2013ja,Shuryak:2013tka,Liu:2015ufa,Liu:2015jsa,Larsen:2015vaa}  on this matter  below in section \ref{ensemble}.  Specifically, we want to pinpoint few crucial elements which  differ between Euclidean expressions and corresponding   formulae  written   in hyperbolic space.   Precisely this difference as we shall argue in section  \ref{sec:hyperbolic} is responsible for the linear in $\kappa$ correction in expression (\ref{FLRW}) describing  the ``strange" energy. 
 
 \subsection{\label{ensemble} The grand canonical ensemble of monopoles with nontrivial holonomy (\ref{polyakov})}
 In the semiclassical approximation the partition function of the gluodynamics is represented by the statistical ensemble of an arbitrary number of interacting monopoles and anti-monopoles of all kinds.  The corresponding picture is well tested in the weakly coupled gauge theory, the so-called ``deformed QCD" where all important  elements such as   the generation of the strange energy, the topological susceptibility, the contact term, etc, have been explicitly computed \cite{Yaffe:2008,Thomas:2011ee,Zhitnitsky:2013hs}.
 The key lesson from that studies is the crucial role of the holonomy (\ref{polyakov}) and non-locality in generation of all these effects. In fact one can argue \cite{Zhitnitsky:2013hs} that the system belongs to  a topologically ordered phase as a result of these non-local effects\footnote{Unfortunately we can not use the ``deformed QCD" model  to address  the question formulated  in the present work. This is because we can not implement $\kappa$ parameter into this model  because the size of $\mathbb{S}^1$  in ``deformed QCD" model \cite{Yaffe:2008}
 must be small    to keep the system  in the weakly coupled regime, while in hyperbolic space  the radius of $\mathbb{S}^1_{1/\kappa}$ must be large as it  is correlated with our large 3d volume $\mathbb{H}^3_{\kappa}$. }.
 
 Similar computations in strongly coupled regime have not been performed yet. Nevertheless, one should expect a very similar behaviour of the strongly coupled ensemble of KvBLL calorons   represented by a  set of their  monopole  constituents as discussed above in section \ref{sec:details}. In other words, we assume that  KvBLL calorons with nontrivial holonomy 
 are responsible for generation     of the strange energy, the topological susceptibility, the contact term, and many other highly nontrivial features of the system. One should emphasize that the corresponding contributions are finite in the infrared (IR) in the large volume limit and small temperature, $(V, \beta)\rightarrow \infty$, in contrast with conventional instanton computations. We do not claim that the semiclassical approximation adopted here  is justified in the strongly coupled regime.
 In fact, it is expected that the fluctuations with typical scales $\sim \Lbar$  change  some numerical estimates.
 However, as we shall argue below the corresponding   fluctuations with scales $\sim\Lbar$ can  not modify the contributions from the far IR   regions  with typical scales $\sim \kappa$,  which is precisely the subject of the present work.

 Therefore, we follow \cite{Diakonov:2004jn,Diakonov:2007nv,Faccioli:2013ja,Shuryak:2013tka,Liu:2015ufa,Liu:2015jsa,Larsen:2015vaa} and represent the grand partition function for the ensemble as follows
 \be
 \label{Z}
{ \cal{Z}}&=& \sum_{K_LK_MK_{\bar L}K_{\bar M}}\prod_{i_L=1}^{K_L} \prod_{i_M=1}^{K_M}\prod_{i_{\bar L}=1}^{K_{\bar L}}\prod_{i_{\bar M}=1}^{K_{\bar M}} \nonumber\\
&\times& \int \frac{ fd^3 x_{i_L}}{K_L!}  \frac{  fd^3 x_{i_M}}{K_M!}  \frac{ fd^3 y_{i_{\bar L}}}{{K_{\bar L}}!}  \frac{ f d^3 y_{i_{\bar M}}}
{{K_{\bar M}}!} \nonumber\\
&\times& e^{-V (x-y)}\cdot \det G[x] \cdot \det G[y],
 \ee
 where $f$ is the ``fugacity" of the monopoles. The fugacity  has been computed in \cite{Diakonov:2004jn}
 in terms of the fundamental parameters of the system such as $\Lbar, \beta$, and holonomy $v$ defined by eq. (\ref{v}). One should emphasize that fugacity $f$  is dimensional parameter which is sensitive to all scales of the problem as it is expressed in terms of zero as well as non-zero modes, see below. The  $3$ dimensional coordinates $x_{i_M}, x_{i_L}$ and $y_{i_{\bar L}}, y_{i_{\bar M}}$  describe the positions of  $M, L$  monopoles and   ${\bar L}, {\bar M} $  anti-monopoles correspondingly. 
   The   $G[x]$ is  a $(K_L+K_M)\times(K_L+K_M)$  and $G[y]$ is  a $(K_{\bar L}+K_{\bar M})\times (K_{\bar L}+K_{\bar M})$ matrices describing the moduli space. Their  explicit form are given in refs. \cite{Diakonov:2004jn,Diakonov:2007nv}.  These matrices represent the  standard   zero mode contributions, and highly sensitive to the IR physics as they  depend on the holonomy  and the long range Coulomb interactions between the monopoles. The corresponding interactions can be traced from the asymptotical behaviour of the monopole's solutions (\ref{M_bc}) and (\ref{L_bc}).  Finally, potential $V (x-y)$ describes the interaction between monopoles and anti-monopoles of the entire ensemble. The corresponding  interactions, along with $G[x]$ and $G[y]$ are also long ranged.  
  These elements  of the partition 
 function are  highly sensitive to the IR behaviour and to the boundary conditions. 
 \exclude{Precisely this Coulomb interaction   will play a key role in our discussions in  section \ref{sec:hyperbolic} where we shall study the same system defined in the hyperbolic space $\mathbb{H}^3_{\kappa}\times \mathbb{S}^1_{\kappa^{-1}}$. }

 There are many subtle points in writing (\ref{Z}) which shall not be discussed here. We refer to the original works 
 for the discussions and references. For the moment we ignore the interaction between the monopoles and antimonopoles, $V(x-y)$. As we argue below in section \ref{sec:fugacity}, the corresponding interaction may change the numerical results, but can  not modify  our main claim on  structure of the correction $\sim \kappa$. 
 
 The only  relevant element for our future studies  is the presence of the long range forces entering $G[x], G[y]$.  The corresponding interactions are effectively cancel  in computation of the free energy due to the total neutrality condition\footnote{\label{neutrality}The consequences on  neutrality condition  has been slightly corrected recently in \cite{Liu:2015ufa,Liu:2015jsa,Larsen:2015vaa}. The correct statement is not that the neutrality condition implies that the total charge   is zero for each given configuration. Rather, the correct statement is that the corresponding expectation value of the charge     vanishes while   charge number density  itself may still fluctuate.}, as argued in  \cite{Diakonov:2004jn,Diakonov:2007nv}.  As a result of neutrality  the number of different  types of monopoles is the same in each given configuration. In other words, the partition function under these assumptions decouples for monopoles and anti-monopoles, ${ \cal{Z}} = { \cal{Z}}_{+}  \cdot{ \cal{Z}}_{-}$, and  takes the following simple form    \cite{Diakonov:2004jn,Diakonov:2007nv}:
 \be
 \label{Z1}
 { \cal{Z}}_{\pm}=\sum_{K_L K_M}\frac{(4\pi fV)^{K_L+K_M}}{K_L! K_M!}\nu^{K_M}(1-\nu)^{K_L} ,
 \ee
where $V$ is the 3-volume of the system. 
The combination $\nu^{K_M}(1-\nu)^{K_L} $ which enters  the partition function (\ref{Z1}) comes form zero mode determinant which itself  is expressed in terms of the classical actions of the constituents (\ref{S}). 
  In the large volume limit the sum 
is saturated by very large $K$ such that the partition function can be evaluated using saddle point approximation. 

To proceed with estimations we first represent $k!$ using the Stirling formula
\be
\label{stirling}
\frac{1}{K!}= e^{-\ln K!}\simeq \frac{1}{\sqrt{2\pi K}}e^{-K\ln K +K}.
\ee
The next step is to replace the sum by the integrals
\be
\label{integral}
&& \sum_{K_L K_M}\rightarrow \int dK\int dQ, \nonumber\\ &&K\equiv K_L+ K_M, ~~ Q\equiv K_M-K_L,
\ee
where $K$ describes the total number of monopoles in a given configuration, while   $Q$ describes the magnetic charge of a given configuration as $M$ and $L$ monopoles have opposite charges according to (\ref{S}). Using saddle point approximation one arrives to the following expression for the partition function in terms of the saddle value $K_0$ which saturates (\ref{Z1})
\be
\label{Z2}
{ \cal{Z}}_{\pm}\simeq  e^{K_0}\cdot \int 
\frac{e^{-\frac{Q^2}{2K_0}}dQ}{\sqrt{2\pi K_0}}, ~~ K_0=  8\pi fV\sqrt{\nu(1-\nu)}. ~~
\ee
 Therefore the final expression for   the partition function and the free energy for the vacuum ground state in this approximation assumes the form
 \be
 \label{Z3}
 { \cal{Z}}_{\pm}\simeq \exp\left[4\pi fV\right], ~~~~ f=\frac{4\pi\Lbar^4}{g^4 T} \nonumber\\ F_{\rm vac}=-T\ln  { \cal{Z}}=-\frac{32\pi^2}{g^4} \Lbar^4 V, 
\ee
 where we substitute the ``confining" value for the holonomy  $\nu=1/2$ which minimizes the free energy. 
 
 Few comments are in order.
  The expectation value $\la Q\ra=0$ obviously vanishes, such that system is  neutral.
  However, the fluctuations of the $Q^2$ do not vanish, in agreement with \cite{Liu:2015ufa,Liu:2015jsa}, but  strongly suppressed for large volume system, as  expected
  \be
  \label{Q}
 \sqrt{ \frac{\la Q^2\ra}{K_0^2}}\simeq \sqrt{\frac{1}{K_0}}\sim \frac{1}{\sqrt{V}} \rightarrow 0.
  \ee  
  The free energy (\ref{Z3}) is finite at zero temperature limit. Furthermore, $\ln  { \cal{Z}}$ is proportional to the  4-volume
  $\sim V/T$ of the system demonstrating the expected extensive scaling at low temperature. One should emphasize that these  computations (based on configurations with nontrivial holonomy) generate the  IR finite and well defined contributions to different observables   expressed in terms of fundamental parameters of the theory, in contrast, for example, with instanton computations. 
   The dynamically generated 
  ``confining" value for the holonomy  $\nu=1/2$  is also  a highly nontrivial phenomenon -- it leads to a proper behaviour for the Polyakov's loop, the Wilson loop and the string tension  \cite{Diakonov:2004jn,Diakonov:2007nv}, see also Appendix \ref{terminology} with few historical and  terminological comments on fractionally  charged constituents, previously emerged   in the literature in different contexts and different systems.  
  
  Furthermore, this ``confining" value for the holonomy  $\nu=1/2$ leads to a consistent resolution of the so-called $U(1)_A$ problem formulated in terms of the topological susceptibility (\ref{top1}) and the $\theta$ dependence   of the ``strange" vacuum energy (\ref{chi}). Indeed, the introduction of the $\theta$ term into the Lagrangian changes the fugacity  for the  monopole $f\rightarrow f e^{i\theta/2}$ and   anti-monopoles $f\rightarrow f e^{-i\theta/2}$. This modification   follows from the fact that the topological charges for monopoles and anti-monopoles assume the magnitude $Q=\pm 1/2$  for ``confining" value of the holonomy  $\nu= 1/2$ as it follows from quantum numbers  for monopoles  (\ref{S}). The  anti-monopoles  assume the opposite sign for the topological charge as they are anti-selfdual solutions.   This modification leads to replacement of expression  (\ref{Z3}) by the following formula  which is valid for $|\theta|\leq \pi$:
   \be
 \label{theta}
   F_{\rm vac} (\theta)= -\frac{32\pi^2}{g^4} \Lbar^4 V \cdot \cos\left(\frac{\theta}{2}\right).
\ee
  The topological susceptibility now can be easily computed by differentiating  (\ref{theta}) twice with respect to $\theta$ with result   
 \be
 \label{theta-1}
  \chi = \left. \frac{1}{V}\frac{\partial^2 F_{\rm vac} (\theta)}{\partial^2\theta} \right|_{\theta=0} =\frac{8\pi^2}{g^4} \Lbar^4 . 
  \ee
  Finally the vacuum energy (\ref{theta}) per unit volume $ F_{\rm vac}/V\sim \Lbar^4 $ is precisely the first term entering the  expression (\ref{FLRW}). It has all the features of the ``strange energy" briefly described in section \ref{contact} in model-independent generic way. 
    The mechanism based on the KvBLL configurations  reviewed above   precisely generates all these required properties. 
   Similar formulae can be easily generalized for arbitrary  number of colours $N$ when $F_{\rm vac} (\theta)\sim N^2  \cos\left(\frac{\theta}{N}\right)$ and  $\chi \sim 1$, which is consistent with conventional resolution of the $U(1)_A$ problem in  large $N$ limit.  
  
  We do not claim to have derived  any new results in the present subsection. Rather,   we just reproduced and explained the known  results \cite{Diakonov:2004jn,Diakonov:2007nv,Liu:2015ufa,Liu:2015jsa} in slightly different and simplified manner in order to analyze   the role of similar  vacuum configurations in cosmological context in next section \ref{sec:hyperbolic}. Furthermore, we do not claim that the corresponding computations in strongly coupled regime are exact.
    In fact, we expect the corrections to be order of one to the fugacity $f$ and  all other numerical coefficients such as (\ref{Z3}), (\ref{theta-1}) discussed above. However, we do not anticipate any drastic  qualitative changes of this framework as a result of  these possible  corrections. In particular, we expect that the   free energy generated by these configurations remains finite in the IR and  demonstrates the extensive   behaviour at low temperature $T\rightarrow 0$ as presented above. Precisely these features will play a crucial role in our arguments on small modification  of this ``strange" vacuum  energy with tiny variation of the background to be considered in section \ref{energy}. 

    \subsection{\label{interpretation}Interpretation. Cosmological context.}
    There are many  important elements related to the computations reviewed in previous sections. In what follows we would like to make only very few comments   which will be relevant for our studies in cosmological context on  IR sensitivity of the system.  
    
1. First of all, the positive sign in (\ref{theta-1}) unambiguously implies  that the corresponding configurations saturating the   topological susceptibility (and related the $\theta$- dependent portion of the vacuum energy) can not be identified with any   propagating degrees of freedom  in accordance with (\ref{G}). Indeed, the   computations reviewed above explicitly  show that the relevant  configurations are the KvBLL calorons 
  with nontrivial holonomy describing the tunnelling events between topologically distinct but physically identical winding states, rather than propagating gluons. All effects are obviously non-analytical  in coupling constant $\sim \exp(-1/g^2)$ and can not be seen in perturbation theory.
  
2. One can view the relevant topological configurations as the 3d   magnetic monopoles  wrapping around time direction. This leads to the non-vanishing holonomy (\ref{polyakov}) and non-vanishing topological charge (\ref{S}) of the constituents  defined in 4d space-time.  In the limit of $T\rightarrow 0$ the ``confining" value $\nu=1/2$ for  the holonomy implies that the parameter $v$ which determines the monopole's mass in conventional 3d theory   tends to zero in this limit, $v=\pi T\rightarrow 0$. At the same time the  monopole   generates  a finite contribution (\ref{Z3})  to the path integral due to its finite 4d action (\ref{S})  resulting  from  very  long path $\sim T^{-1}$  in  time direction. It would be misleading to interpret the confinement and other features (discussed above in section \ref{ensemble})  generated by  these configurations as the condensation of the monopoles. It would be more appropriate to use term ``percolation"
  as the configurations described above do  correlate at arbitrary large distances, but they obviously do  not   form a  ``condensate" in conventional condensed matter terminology. 
  
3. In the cosmological context such configurations are highly unusual objects: they obviously describe the non-local physics 
  as the holonomy (\ref{polyakov}) is a nonlocal object. Indeed,  the holonomy  defines the dynamics  along the entire history  of evolution of the system in the given confined phase: from the very beginning to the very end. There is no contradictions with causality    in the system as there is no any   physical degrees of freedom   to propagate along this   path at $\beta\rightarrow\infty$, see item 1. above. Indeed,  this entire gauge configuration is a mere  saddle point in Euclidean (imaginary time) path integral computation which describes the instantaneous tunnelling event, rather than propagation of  a physical  degree of freedom capable to carry an information/signal.  
  
 4. Further to this point,   the extensive property of free energy $\beta F_{\rm vac} \sim V/T$ at $\beta\rightarrow \infty$ is a highly nontrivial  phenomenon in this framework as the 4-volume $V/T$ appears in this  description due to few important steps. First,   one should regularize  the moduli space by cutting off an  each given configuration in the IR. Secondly, one should  sum over all configurations  (\ref{Z1}) by  using saddle point approximation in  large volume limit, which eventually  leads to (\ref{Z3}). This ``emergent" extensive property  is drastically different from conventional approaches to cosmology  when  the free energy is determined by the  Lagrangian density $L[\Phi]$ integrated over the 4-volume $\int  d^4x$. In this last case the 
   extensive property   is a trivial manifestation of  the system formulated from the very beginning  
 in terms of the local field $\Phi(x)$.      It shows one more time that generation of the ``strange" energy (\ref{FLRW}) is highly non-local non-perturbative effect when the  volume of the system could be very large, but still finite to proceed with computations (\ref{Z3}) in this approach.  
   Essentially, the finite local energy density of the system (\ref{Z3}) in this framework is determined by the entire time evolution  $\beta\rightarrow \infty$ in confined phase, which is obviously a non-local procedure. Still, it does not contradict  the causality, see item 3 above.  
     
  5. Last, but not least. All these highly nontrivial non-local features listed above emerge only   at $T<T_c$ when the configurations with nontrivial holonomy (\ref{polyakov})  start to play the dominant role in the dynamics.  At   high temperatures the contribution of the configurations with nontrivial holonomy  can be completely ignored  as they do not contribute to the partition function in thermodynamical limit. This property of drastic variation of ``strange" energy (\ref{FLRW}) with temperature around $T_c$ may play an important role in cosmological context as we discuss in next section.

\section{\label{sec:hyperbolic} Non-trivial holonomy and hyperbolic space 
$\mathbb{H}^3_{\kappa}\times \mathbb{S}^1_{\kappa^{-1}}$}
The main goal of this section is to generalize the results presented above in sections \ref{sec:details}, \ref{ensemble} to hyperbolic space 
$\mathbb{H}^3_{\kappa}\times \mathbb{S}^1_{\kappa^{-1}}$  to argue that the correction to the free energy (\ref{Z3}) are linearly proportional to $\kappa$ at small $\kappa\rightarrow 0$. In this limit it is quite obvious that all features listed  in section \ref{interpretation}   on nature of the ``strange energy", including  its non-local nature,   remain the same in this limit $\kappa\rightarrow 0$ as the system is almost 4d Euclidean space, with very tiny deviations $\sim \kappa$ which we wish to recover. The corresponding linear dependence on $\kappa$ would strongly support our conjecture that the correction in eq. (\ref{FLRW}) are linearly proportional to the Hubble constant\footnote{see footnote \ref{H} with some clarification on terminology.},   which dynamically drugs  our Universe to the de Sitter state  as Friedman equation (\ref{friedman-infl}) suggests. 

We discuss the relevant gauge configurations in hyperbolic space in subsection \ref{monopoles}, while the grand canonical ensemble of such hyperbolic monopoles will be studied in subsection \ref{ensemble-hyperbolic} where   we discuss the crucial distinctions between hyperbolic and Euclidean monopoles. Precisely this difference    eventually leads to  a tiny $\sim \kappa$ variation from the Euclidean results. 
The corresponding deviation is expressed in terms of the fugacity in section \ref{sec:fugacity}, which ultimately  leads to slight modification of the vacuum energy. We list  some profound cosmological consequences of this modification of the vacuum energy with background  in section \ref{energy}. 

\subsection{\label{monopoles} Holonomy and monopoles  in hyperbolic space} 
The construction of the monopoles on hyperbolic space $\mathbb{H}^3_{\kappa}$ has  been known since \cite{Atiyah,Chakrabarti:1984sm, Nash}. Furthermore, many other topological objects, including  calorons, instantons, vortices,  skyrmions   have been constructed on hyperbolic space \cite{Garland:1989cd,Manton:1990gr,Atiyah:2004nh,Harland:2007cq,Sutcliffe:2012pu,Manton:2012xn}. An   important technical element which was used in  these constructions is   the conformal equivalence of  $\mathbb{R}^4$ and $\mathbb{H}^3_{\kappa}\times \mathbb{S}^1_{\kappa^{-1}}$. 
Indeed, this equivalence can be explicitly checked  by introduction toroidal coordinates $(\rho,\theta,\phi,\chi )$ on $\mathbb{R}^4$ as follows:
\be
\label{coordinates}
x_{\mu}&=& \left(x_1, x_2, x_3, x_4 \right)\\
x_1&=& \frac{1}{\cosh (\kappa\rho)+\cos\chi} \sinh (\kappa\rho)\sin\theta\cos\phi ,\nonumber\\
x_2 &=&  \frac{1}{\cosh (\kappa\rho)+\cos\chi} \sinh (\kappa\rho)\sin\theta\sin\phi , \nonumber\\
x_3 &=&  \frac{1}{\cosh (\kappa\rho)+\cos\chi}\sinh (\kappa\rho)\cos\theta ,\nonumber\\
x_4 &=&\frac{1}{\cosh (\kappa\rho)+\cos\chi} \sin\chi . \nonumber
 \ee
It is then easy   to check that the metric on $\mathbb{R}^4$ becomes
\be
\label{metric}
ds^2 (\mathbb{R}^4)\equiv dx_{\mu}dx^{\mu}= \frac{ ds^2 (\mathbb{H}^3_{\kappa})+\kappa^{-2} d\chi^2}{(\cosh (\kappa\rho)+\cos\chi)^2}, 
\ee
where $ds^2 (\mathbb{H}^3_{\kappa})$  is the metric on hyperbolic 3-space with spherical coordinates $(\rho,\theta, \phi)$:
\be
\label{metric1}
ds^2 (\mathbb{H}^3_{\kappa})=d\rho^2+\frac{\sinh^2(\kappa\rho)}{\kappa^2} \left(d\theta^2+\sin^2\theta d\phi^2\right).
\ee
The holonomy in terms of these variables is computed along the circle $ \mathbb{S}^1_{\kappa^{-1}}$ parameterized by $\chi$, that is 
\be
\label{U-hyperbolic}
U ={\cal{P}}\exp\left(\frac{i}{\kappa}\int_0^{2\pi} {d\chi} A_{\chi}\right),
\ee
where $A_{\chi}$ is the component of the gauge potential associated with coordinate $\chi$, and it plays the same role as the $A_4$, similar to the construction in Euclidean space (\ref{M}). In both cases, the $A_4$ and $A_{\chi}$ assume  non-vanishing expectation values, and play the same role as the Higgs field $\Phi$ in adjoint representation, as explained after eq. (\ref{v}), see also   footnote \ref{normalization} on our normalization.  Formula (\ref{U-hyperbolic}) plays the same role as equations (\ref{holonomy}), (\ref{polyakov}), while $d\chi/\kappa$ in eq. (\ref{U-hyperbolic}) plays the role of $dx_4$ in evaluation  of the holonomy computed along the circle $ \mathbb{S}^1$ according to (\ref{v}). At large $\rho\rightarrow \infty$ 
the $A_{\chi}$ approaches a non-vanishing constant value, similar to parameter $v$ in Euclidean space (\ref{M}).  

With these remarks in mind    the explicit form for   the  Bogomolny- Prasad- Sommerfeld (BPS)  monopole  in hyperbolic space $\mathbb{H}^3_{\kappa}$  in the unitary gauge can be written as follows \cite{Chakrabarti:1984sm,Nash}:
\be
\label{M-hyperbolic}
A_{\chi}^{M}(\rho)=\left[ {C\kappa} \coth (C\kappa\rho)-{\kappa}\coth (\kappa\rho)\right] \frac{\tau^3}{2},
\ee
where parameter $C$ takes any value greater than 1. In formula (\ref{M-hyperbolic}) we limited ourselves by 
  writing    down  only the $A_{\chi}^M (\rho)$-component, which  defines the boundary conditions at large $\rho$.    We  coin  this   solution  as $M$ monopoles in order to be consistent with the terminology  introduced for the Euclidean counterparts (\ref{M}). 

While many topological objects, including calorons with trivial holonomy \cite{Harland:2007cq} have been   constructed in hyperbolic space, as we already mentioned, an explicit  construction of the calorons with 
nontrivial holonomy,  which is analogous to the KvBLL solutions   \cite{Kraan:1998pm,Lee:1998bb}, has not been constructed yet.
In what follows we shall assume that such solutions do exist, though we do not need their explicit form in our future discussions.  Important point is that if such configurations  exist   than  they must exhibit  the same   features which    the KvBLL solutions demonstrate. Namely they could  be viewed as a set of $2$ different types of monopoles for $SU(2)$ gauge group. The first type is precisely the M-monopole (\ref{M-hyperbolic}) discussed above, while the second one, the $L$ monopole can be constructed as described    in section \ref{sec:details} for the Euclidean counterpart. We shall return to this construction later in the text, but first, we want to understand the physical meaning of the parameter $C$ entering eq. (\ref{M-hyperbolic}) by analyzing the limit $\kappa\rightarrow 0$ when the Euclidean monopole (\ref{M}) is recovered.

To recover the Euclidean monopole solution one should take the limit  $C\rightarrow \infty$ along with $\kappa\rightarrow 0$ 
with combination $C\kappa$ being fixed to be equal $v$. In this limit $A_{\chi}^{M}(\rho)$ becomes 
\be
\label{limits}
A_{\chi}^{M}(\rho)=\left[ v \coth (v\rho)- \frac{1}{\rho}\right] \frac{\tau^3}{2}, ~~~~~~  C\kappa\equiv v 
\ee
where $\rho$ should be identified with $r$ in the Euclidean space. 
Expression (\ref{limits})  exactly coincides with  (\ref{M})   and its asymptotic behaviour (\ref{M_bc}) in the unitary gauge.
 
We are now in position to discuss the asymptotical behaviour of (\ref{M-hyperbolic}) and the holonomy (\ref{U-hyperbolic})  for finite $\kappa$. Taking $\rho\rightarrow \infty$ one arrives to the following expressions  for $A_{\chi}^{M}(\rho)$ and   holonomy
\be
\label{holonomy-hyperbolic}
A_{\chi}^{M}(\rho\rightarrow \infty)\rightarrow \left[ \kappa (C-1)+ {\cal{O}} (e^{-\rho}) \right] \frac{\tau^3}{2}, \nonumber \\
\frac{1}{2}\Tr U(\rho\rightarrow \infty)=\cos \pi \nu, ~~~ \nu\equiv (C-1), 
\ee
 where we introduced parameter $\nu$ expressed in terms of original parameter $C$. It plays the same role as parameter $\nu$ discussed in the Euclidean construction (\ref{holonomy}), (\ref{S}). The crucial observation here   is that the asymptotic formula for $A_{\chi}^{M}(\rho)$ does not contain a long range Coulomb interaction $\rho^{-1}$ in contrast with  its  Euclidean counterpart (\ref{M_bc}). Instead, there is an exponentially suppressed correction $\sim \exp{(-\kappa\rho)}$ in formula (\ref{holonomy-hyperbolic}).   One can interpret such drastic changes in behaviour of the solution as a result of screening  of the magnetic field by the curvature in the hyperbolic space. In terms of the new parameter  $\nu$ the solution (\ref{M-hyperbolic}) assumes the form
 \be
 \label{M-hyperbolic-nu}
 A_{\chi}^{M}(\rho)=\Big(({\nu}+1) \coth \left[({\nu}+1)\kappa\rho\right]-\coth \kappa\rho\Big) \frac{\kappa\tau^3}{2}. ~~
\ee

Our next step is to recover  the $L$ monopole assuming that KvBLL caloron with nontrivial holonomy  in the hyperbolic space exists,   similar to construction \cite{Kraan:1998pm,Lee:1998bb} in the Euclidean space.  We follow the same steps to reconstruct their properties presented  in section  \ref{sec:details} for the Euclidean monopolies.  The first step is to replace $\nu\rightarrow (1-\nu)$. The second step is to make a ``large"  gauge transformation which  assumes the following form in hyperbolic variables
\be
\label{U-large}
U_{\rm large} (\chi)=\exp\left(i\frac{\tau_3}{2}\chi\right).
\ee
As we already mentioned  the ``large" gauge transformations should be treated differently from ``small"  (proper) gauge transformations because any two field configurations related by ``large"  gauge transformation do not belong to the same gauge orbit. Nevertheless, the transformation (\ref{U-large})  preserves the periodic boundary conditions for the fields in the adjoint representation   because
$U_{\rm large} (\chi=0)=- U_{\rm large} (\chi=2\pi)$. The final step is to perform the discrete transformation    $U_{\rm reflection}=\exp(i\tau^2{\pi}/{2})$ to restore the original boundary conditions (\ref{holonomy-hyperbolic}). The resulting configuration is the hyperbolic $L$- monopole:
 \be
\label{L-hyperbolic-nu}
A_{\chi}^{L}(\rho)=\Big(1- (\bar{\nu}+1) \coth \left[(\bar{\nu}+1)\kappa\rho\right]+\coth \kappa\rho\Big) \frac{\kappa\tau^3}{2},~~~~~
 \ee
 where we introduced $\bar{\nu}\equiv (1-\nu)$ for convenience.  One should emphasize that electric and magnetic fields of the 
 the $L$ -monopoles do depend on  $\chi$ variable   as a result of $\chi$ dependent 
 ``large" gauge transformation (\ref{U-large}).  This  is  analogous to $L$ -monopole solution in Euclidean space (\ref{L_bc})
 which is a time dependent configuration,   and cease to  exist in static 3d space. 
 
The classical action, topological and magnetic charges of the $M$ and $L$ constituents are determined by the boundary conditions
(\ref{M-hyperbolic-nu}) and (\ref{L-hyperbolic-nu}) at large $\rho\rightarrow\infty$, similar to the Euclidean counterparts (\ref{S}). 
The corresponding parameters   $S, Q, q$ obviously assume the same values (\ref{S}) when expressed in terms of $\nu$. These dimensionless parameters obviously can not depend on dimensional parameter $\kappa$, including $\kappa\rightarrow 0$ limit. In fact, the corresponding formula relating $Q$ and $q$ for BPS $M$- type monopole (\ref{M-hyperbolic-nu}), which identically coincides with the Euclidean expression,  was derived in hyperbolic space in the original work \cite{Atiyah}, while the relation between $S$ and $Q$ is a direct consequence of self -duality of Yang-Mills equations. The monopole's mass  (the total energy of the configuration), being a dimensional parameter, does depend on $\kappa$. Mass $m$ satisfies an obvious relation $S=m\frac{2\pi}{\kappa}$, similar to its Euclidean counterpart, see text after eq. (\ref{S}).   Explicitly, in terms of the original parameter $C$ it is  given  by   $m=\frac{4\pi}{g^2}\kappa (C-1)$, which reduces to its Euclidean form $m=\frac{4\pi}{g^2}v$ when one takes the corresponding limit (\ref{limits}). 

\subsection{\label{ensemble-hyperbolic} Grand partition function for  monopoles  in the hyperbolic space} 
 With our main assumption that the calorons with nontrivial holonomy exist in hyperbolic space, similar to Euclidean KvBLL construction \cite{Kraan:1998pm,Lee:1998bb}, one should expect  that   the corresponding grand partition function ${\bar{{ \cal{Z}}}}$ has  the following form,   which is analogous to expression  (\ref{Z}) discussed in previous section:  
\be
 \label{Z-hyperbolic}
{\bar{{ \cal{Z}}}}&=& \sum_{K_LK_MK_{\bar L}K_{\bar M}}\prod_{i_L=1}^{K_L} \prod_{i_M=1}^{K_M}\prod_{i_{\bar L}=1}^{K_{\bar L}}\prod_{i_{\bar M}=1}^{K_{\bar M}} \nonumber\\
&\times& \int_{\mathbb{H}^3_{\kappa}} \frac{ \bar{f}\sqrt{g}d^3 \bar{x}_{i_L}}{K_L!}  \frac{\bar{f}\sqrt{g}d^3 \bar{x}_{i_M}}{K_M!}  \frac{\bar{f}\sqrt{g}d^3 \bar{y}_{i_{\bar L}}}{{K_{\bar L}}!}  \frac{\bar{f}\sqrt{g}d^3 \bar{y}_{i_{\bar M}}}
{{K_{\bar M}}!} \nonumber\\
&\times& e^{-V (\bar{x}-\bar{y})}\cdot \det G[\bar{x}] \cdot \det G[\bar{y}],
 \ee
 where $\bar{f}$ is the ``fugacity" of the hyperbolic monopoles. The corresponding dimensional parameter is highly sensitive 
 to many details of monopole's structure and their interactions with other monopoles.  It is obviously different from its cousin fugacity $f$  computed in the Euclidean space as discussed in section \ref{ensemble}. We shall estimate  the  corrections to 
 $\bar{f}$ later in section \ref{sec:fugacity} where we argue that  the difference  $(\bar{f}- f)\sim \kappa$ is  linear in $\kappa$ at small $\kappa\rightarrow 0$ for configurations with nontrivial holonomy. 
   
  The next item to   discuss from formula (\ref{Z-hyperbolic}) is the  $3-$ dimensional coordinates $\bar{x}_{i_M}, \bar{x}_{i_L}$ and $\bar{y}_{i_{\bar L}}, \bar{y}_{i_{\bar M}}$. They   describe the positions of  $M, L$  monopoles and   ${\bar L}, {\bar M} $  anti-monopoles correspondingly.    These coordinates  play the same role as in formula (\ref{Z}) with the only difference is that the distance between constituents is computed using the metric (\ref{metric1}).
   The   $G[\bar{x}]$ is  a $(K_L+K_M)\times(K_L+K_M)$  and $G[\bar{y}]$ is  a $(K_{\bar L}+K_{\bar M})\times (K_{\bar L}+K_{\bar M})$ matrices describing the moduli space. The difference with corresponding Euclidean expressions  is that the   behaviour at large distances is not the Coulomb like, but rather the exponentially suppressed as we already mentioned (\ref{holonomy-hyperbolic}). This is because the matrices $G[\bar{x}]$ and $G[\bar{y}]$ are computed from  the corresponding zero modes in the background of the monopoles. At the same time, the zero modes, as usual,  are fixed by the corresponding  classical solutions.  Therefore, the  asymptotical behaviour of  classical solutions (\ref{M-hyperbolic-nu})
   and (\ref{L-hyperbolic-nu}) dictates  the behaviour of $G[\bar{x}]$ and $G[\bar{y}]$.  
   In both cases the computations of the free energy is reduced to expression (\ref{Z1})    as a result of neutrality condition discussed  in   section \ref{ensemble}.     
  
   Final item to discuss is  the factor $\sqrt{g}$ which accounts for the curvature of  the hyperbolic space $\mathbb{H}^3_{\kappa}$.  It enters along with the spatial volume of the system where computations are being performed. We put the system into a large volume of radius $R$ such that the volume is
   \be
   \label{volume}
   V=\frac{4\pi}{2\kappa^3} \left(\frac{\sinh 2\kappa R}{2}-\kappa R \right), ~~\sqrt{g}=\frac{\sin\theta\sinh^2\kappa \rho}{\kappa^2}. ~~~
   \ee
   It reduces to the Euclidean  expression $V=\frac{4\pi R^3}{3}$ in the limit of small $\kappa$ with the corrections  of order ${\cal{O}}(\kappa^2)$ which are consistently neglected in the present work. Therefore, by repeating  all the steps leading to formula (\ref{Z3}) we arrive to the following expression for the partition function and the free energy in hyperbolic space
    \be
 \label{Z3-hyperbolic}
{\bar{{ \cal{Z}}}}_{\pm}\simeq \exp\left[4\pi \bar{f}V\right], ~~~~ \bar{F}_{\rm vac}=-\frac{\kappa}{2\pi}\ln {\bar{{ \cal{Z}}}}, 
\ee
where we identify $ \mathbb{S}^1$ from $\mathbb{R}^3 \times \mathbb{S}^1$ geometry (reviewed in section \ref{review})  with circle $ \mathbb{S}^1_{\kappa^{-1}}$  from  $ \mathbb{H}^3_{\kappa}\times \mathbb{S}^1_{\kappa^{-1}}$ geometry presented in section \ref{monopoles}. In other words, we identify
\be
\label{identification}
T\equiv\frac{1}{\beta}=\frac{\kappa}{2\pi}, 
\ee
such that the corresponding expressions for holonomy (\ref{holonomy}) and (\ref{U-hyperbolic}) coincide. With this identification the corresponding formulae (\ref{holonomy1}) and (\ref{holonomy-hyperbolic}) in terms of dimensionless parameter $\nu$ also coincide.  In formula (\ref{Z3-hyperbolic}) we substitute the confining value for holonomy  $\nu=1/2$ as it has been done in the Euclidean space. This is  because
  the free energy is minimized at  $\nu=1/2$ irrespectively to the value of the fugacity, which is indeed different for two different geometries.  To conclude:  the only difference    between 
${\bar{{ \cal{Z}}}}$ and $ { \cal{Z}}$ describing the system  on   $ \mathbb{H}^3_{\kappa}\times \mathbb{S}^1_{\kappa^{-1}}$ and $\mathbb{R}^3 \times \mathbb{S}^1$  geometries  correspondingly  is that the fugacities   in these two systems assume slightly  different values at small $\kappa\rightarrow 0$,  which is the subject of  the next subsection. 

\subsection{\label{sec:fugacity} Monopole's fugacity  }  
We start our analysis with explanations on how the basic dimensional parameter, the fugacity, emerges  in the system.
This $x$ independent dimensional parameter $f$ effectively  determines  the dynamics of the system. This parameter essentially represents the density of the monopoles  in the system.   
    The classical action, the zero   and  nonzero mode  contributions  lead to the following expression for   the monopole's fugacity $f$   in terms of the fundamental parameters of the theory  \cite{Diakonov:2004jn,Diakonov:2007nv}:
\be
\label{fugacity}
f^2&=&\left[\frac{4\pi\beta\Lbar^4}{g^4} \right]^2\cdot c \\
c&=& \langle \frac{[1+2\pi \nu\bar{\nu}\frac{r_{12}}{\beta}]}{(\Lbar~ r_{12})^{^{2/3}}}[1+2\pi \nu \frac{r_{12}}{\beta} ]^{\frac{8}{3}\nu-1} [1+2\pi \bar{\nu} \frac{r_{12}}{\beta} ]^{\frac{8}{3}\bar{\nu}-1}\rangle , \nonumber
\ee 
where brackets $\langle ...\rangle$ imply averaging over separation $r_{12}$ between $M$ and $L$ monopoles in ensemble  (\ref{Z}) such as $f$  is  $x$-independent as it should.  

Few comments are in order. Each KvBLL caloron is represented by  the $L$ and $M$ monopoles and accompanied by 8 zero modes.   It  explains the  major  dimensional factor  in eq. (\ref{fugacity}), including $\Lbar^8$. The remaining numerical    dimensionless factor  $``c"$ entering  (\ref{fugacity}) is order  of 1. It includes factor $(\Lbar)^{-{2}/{3}}$ which can be easily restored from the renormalization group analysis which requires that $\Lbar$ enters (\ref{fugacity}) with power $(\Lbar)^{{22}/{3}}$. Subsequently, this factor $(\Lbar)^{-{2}/{3}}$  must be accompanied by a dimensional parameter, which at small temperatures could be nothing else but the  separation distance $r_{12}^{-2/3}$ between the monopoles.  In  estimates  of refs.\cite{Diakonov:2007nv} the numerical coefficient $c$ is assumed to be one, which precisely corresponds to the expression for $f$ given in  eq. (\ref{Z3}). 

Next factor which can be easily explained is the first term in numerator,  $[1+2\pi \nu\bar{\nu}\frac{r_{12}}{\beta}]$. 
This term has been originally computed in \cite{Kraan:1998pm} and reproduced in \cite{Diakonov:2004jn,Diakonov:2007nv}. It is originated from  the zero mode  determinant. The crucial point here is that the algebraic  dependence on $r_{12}$   emerges  as a result of  long range Coulomb terms in the classical solutions (\ref{M_bc}) and (\ref{L_bc}). This is because  the  zero mode structure is   unambiguously fixed  by the classical solutions with the corresponding Coulomb terms. Another important element is  that this terms is proportional to the  holonomy $\nu\bar{\nu}$. 
It implies that this term will  not be generated for  the configurations with trivial holonomy. 

The nature of next two terms in numerator in eq. (\ref{fugacity}) is much harder to explain because they are originated from the contributions  of the nonzero modes.  
The only comment we would like to make here is that these terms are also proportional to the holonomy, and can not be generated by the configurations with trivial holonomy.  

Now we are in position to estimate the difference between the fugacity generated by monopoles in the Euclidean space versus hyperbolic monopoles at small $\kappa\rightarrow 0$. The estimation  is convenient to represent in the following form
\be
\label{fugacity1}
\frac{f}{\bar{f}}\simeq (1+\Delta_{\rm zm})\cdot (1+ \Delta_{\rm nzm}), ~~  \Delta_{\rm zm}\simeq  \frac{\nu\bar{\nu}}{2}\frac{\kappa}{\Lbar},
\ee
where factors $\Delta_{\rm zm}$ and $ \Delta_{\rm nzm}$ describe the corrections   due to the zero and non-zero modes correspondingly. The correction factor  $ \Delta_{\rm zm}$ comes from the first term in numerator (\ref{fugacity}), where we estimate $\la r_{12}\ra \sim \Lbar^{-1}$ as the only dimensional parameter in  the system at small temperature. We also  expressed the result in terms of $\kappa$ rather than $\beta$ according to  identification (\ref{identification}). The crucial point in our estimate $ \Delta_{\rm zm}$  is that a similar correction  in hyperbolic space is absent  
as the corresponding classical solutions (\ref{M-hyperbolic-nu}) and  (\ref{L-hyperbolic-nu}) have exponentially suppressed asymptotic at large distances, in contrast with the Euclidean counterpart.  Therefore, a term $\sim r_{12}$ can not be  generated in hyperbolic space, in contrast with the Euclidean case (\ref{fugacity}).  

Unfortunately, a similar unambiguous conclusion   can not be reached   regarding the nonzero mode contribution  $ \Delta_{\rm nzm}$. This is  because both, the Euclidean as well as hyperbolic monopoles may generate such contributions proportional to $\kappa\cdot  r_{12}$, which eventually produce a desired correction $\sim {\kappa}/{\Lbar}$. While the Euclidean expression is known and is represented by two factors in numerator in eq. (\ref{fugacity}), a similar expression in hyperbolic space is not known  simply because  an explicit construction of the hyperbolic calorons with nontrivial holonomy, similar to KvBLL solution,  is yet  unknown. Therefore,   for the numerical estimates in what follows  we set $ \Delta_{\rm nzm}=0$.
 
Few comment are   in order.  
First,  we want to argue that an unknown    $ \Delta_{\rm nzm}$ correction  can not exactly cancel  the   computed $\Delta_{\rm zm}$  term (\ref{fugacity1}).    Indeed, the $\Delta_{\rm zm}$ structure  is precisely fixed by the  structure of the $SU(N)$ gauge group with $4N$ zero modes, while $\Delta_{\rm nzm}$  varies and  depends, in particular,  on presence of the    matter fields, and other details of   the system.   In other words, a possible cancellation, if ever occurs,   can not be a universal phenomenon. Therefore, we use $\Delta_{\rm zm}$ and disregard  $\Delta_{\rm nzm}$  as    our order of magnitude estimate for ratio (\ref{fugacity1}) .  It explicitly exhibits the linear in $\kappa$ correction to the fugacity, which   is the  main result  of the present work. This correction can be  only  generated by the   configurations with  nontrivial holonomy. In particular, conventional instantons and calorons with trivial holonomy may only generate the higher order corrections $\sim {\cal{O}}(\kappa^2)$ and do not contribute to the linear term (\ref{fugacity1}).

We should emphasize that estimate (\ref{fugacity1}) was derived under  assumption that  the   interaction term $V (\bar{x}-\bar{y})$ in formula (\ref{Z-hyperbolic}) vanishes. It corresponds to the ensemble  containing exclusively  the  monopoles (or anti-monopoles). Only  in this case the partition function is exactly reduced to simple form  (\ref{Z1}) as a result of neutrality condition as argued in \cite{Diakonov:2004jn,Diakonov:2007nv}. In reality the interaction plays crucial numerical role at finite temperature as shown  in \cite{Liu:2015ufa,Liu:2015jsa,Larsen:2015vaa}. However, our main claim is that the linear correction (\ref{fugacity1}) may receive large  numerical corrections, but  it can not be exactly cancelled as a result of   unaccounted  interaction term $V (\bar{x}-\bar{y})$ in formula (\ref{Z-hyperbolic}). The basic argument behind this claim  is the same one as presented above and based on observation that  the   interaction  $V (\bar{x}-\bar{y})$ is highly sensitive to the matter content of the theory (the number of flavours and its masses in the system), while a $ \Delta_{\rm zm}$ in formula (\ref{fugacity1}) is not sensitive to these modifications.  The interaction $V (\bar{x}-\bar{y})$  may change a numerical coefficient in the estimate $\la r_{12}\ra \sim \Lbar^{-1}$ which enters (\ref{fugacity1}), but can not completely destroy this term.    Therefore, our main claim (that the linear correction  (\ref{fugacity1}) will be generated)  holds irrespectively to  any type of monopole-anti monopole interactions  $V (\bar{x}-\bar{y})$.

\subsection{\label{energy} Linear corrections $\sim \kappa$ to the vacuum energy}
The result (\ref{fugacity1})  for monopole's fugacity can be translated into the statement on variation of the vacuum energy density in the bulk of space-time 
with a tiny variation of the background. Indeed, according to (\ref{Z3}), (\ref{Z3-hyperbolic}) and (\ref{fugacity1}) the relevant ratio for the vacuum energies at $\kappa\rightarrow 0$ for two different geometries  can be represented as follows
\be
\label{ratio}
\frac{E_{\rm vac}[ \mathbb{H}^3_{\kappa}\times \mathbb{S}^1_{\kappa^{-1}}]}{E_{\rm vac}[ \mathbb{R}^3 \times \mathbb{S}^1]}
\simeq\frac{\bar{f}}{f}\simeq  \left(1-  \frac{\nu\bar{\nu}}{2}\cdot \frac{\kappa}{\Lbar} \right).
\ee
The same result can be represented in more conventional form 
\be
\label{vacuum_energy}
&&E_{\rm vac}[ \mathbb{H}^3_{\kappa}\times \mathbb{S}^1_{\kappa^{-1}}]\simeq  - \Lbar^4  \left(1-  \frac{\nu\bar{\nu}}{2}\cdot \frac{\kappa}{\Lbar} \right)\nonumber\\
&&\simeq -  \Lbar^4+\kappa\cdot \Lbar^3\frac{\nu\bar{\nu}}{2}, 
\ee
where we omitted all irrelevant numerical factors in expression for the vacuum energy in Euclidean space, but kept  the relevant sign minus $(-)$ in front, which is well known feature of QCD. Our final formula (\ref{vacuum_energy}) is a precise analog (in a simplified model) for the vacuum energy (\ref{FLRW}) conjectured for the de Sitter space.  
As we emphasized in the Introduction, the significance of the linear correction in eq.  (\ref{FLRW}) is that the Friedman equation (\ref{friedman-infl})  
unambiguously predicts a non trivial solution with constant $H_0$ if   
 the subtraction procedure is adopted  as discussed in Introduction. 
 The constant solution $H_0$ automatically corresponds to a desired de Sitter behaviour (\ref{a}), which might be relevant for the early Universe during the inflationary epoch, and  in present epoch for description of the dark energy. 
Few comments are in order:

1. The difference between two geometries, $ \mathbb{H}^3_{\kappa}\times \mathbb{S}^1_{\kappa^{-1}}$ and $\mathbb{R}^3 \times \mathbb{S}^1$ when the sizes of $ \mathbb{S}^1_{\kappa^{-1}}$ and $ \mathbb{S}^1$ are identically coincide according to (\ref{identification}) is the small curvature $\sim \kappa^2$ of the hyperbolic space. According to conventional arguments on locality as discussed in Introduction it unambiguously suggests that all corrections must be proportional to the even powers $\kappa^{2n}$. However, we obviously observe a liner correction (\ref{vacuum_energy}) in explicit computations. 

2. This linear correction $\sim \kappa$ is generated by the configurations with nontrivial holonomy, which itself is a non-local, but gauge invariant  operator. Therefore, the standard arguments on locality, reviewed in Introduction  are badly violated by such configurations.  
One can  see from eq. (\ref{vacuum_energy}) that the linear  correction $\sim \kappa$ is explicitly proportional to the holonomy $\nu$, which is the gauge invariant observable, not reducible to the curvature.  In other words, this correction is generated by non-local configurations, and can not be expressed in terms of local curvature $\sim \kappa^2$. 

3. All effects discussed in the present work are non-analytical in coupling constant $\sim \exp(-1/g^2)$ and can not be seen in perturbation theory.

4. The result  (\ref{vacuum_energy}) is consistent with the previous analysis  in weakly coupled ``deformed QCD" model where one can study the sensitivity of the vacuum energy  to the very large distances by putting the system into the box of size $\mathbb{L}$. It turns out \cite{Thomas:2012ib} that  the corrections to the vacuum energy are linear in inverse size $\sim \mathbb{L}^{-1}$. This model is very similar in all respects to the system studied in the present work because the vacuum energy in ``deformed QCD" model    is also saturated by the monopoles with nontrivial holonomy.   At the same time the conventional instantons with trivial holonomy produce only quadratic corrections $\sim \mathbb{L}^{-2}$ as noticed in  \cite{Thomas:2012ib}. 

5.
The generation of the linear correction $\sim \kappa$ is also consistent with computations of ref. \cite{Zhitnitsky:2013pna} in ``deformed QCD" model 
where analysis was  performed in terms of auxiliary topological non-propagating field. In that  computations the root of the phenomenon is the presence of the  non-trivial holonomy and long range monopole's field which eventually is  responsible for generation of the linear correction. 

6.  The linear correction observed in our work is also consistent with the lattice simulations \cite{Holdom:2010ak} when one studies the dependence of the vacuum energy on the size of the system.  

7. Our results are also consistent with the lattice simulations  \cite{Yamamoto:2014vda} when the author studies the rate of particle production in the de Sitter background. The rate turns out to be linearly proportional to the Hubble constant $\sim H$, rather than $H^2$. Our comment here is that the rate of the particle production in quantum field theory in general is determined by the imaginary part of the stress tensor, Im$ [T_{\mu}^{\nu}]$, while the vacuum energy is related to the real part of the stress tensor, Re$[T_{\mu}^{\nu}]$.  Analyticity suggests that both components must have the same corrections  on $H$ at small $H$. Therefore, the lattice measurements 
 \cite{Yamamoto:2014vda} of the linear dependence on $H$ strongly suggest that the vacuum energy (which is determined by the real part of the same stress tensor) must also exhibit the same linear $\sim H$ correction. The corresponding lattice computations of the $\theta$ dependent portion of the vacuum energy and topological susceptibility in time dependent background are possible in principle, but technically much more involved than the  analysis performed in ref. 
 \cite{Yamamoto:2014vda}. 

8. Last but not least. The sign for the difference $\Delta E_{\rm vac}\equiv E [ \mathbb{H}^3_{\kappa}\times \mathbb{S}^1_{\kappa^{-1}}]-E[g_{\mu\nu}=\delta_{\mu\nu}]$ is positive as one can see from eq. (\ref{vacuum_energy}). It corresponds to the positive sign for the cosmological constant
(dark energy) in cosmological context.  

There is a  fundamental difference in signs with conventional Casimir effect when  the corresponding subtraction procedure    typically leads to the negative, rather than positive, sign for the vacuum energy.   This difference is due to the fact that the conventional Casimir vacuum energy is generated by  the fluctuations of the physical propagating photons.
It is drastically different from the   the vacuum energy computed in the present work  when it is generated  by the tunnelling transitions between different topological sectors. As we explained in section \ref{contact} the corresponding vacuum energy can not be expressed in terms of any propagating degrees of freedom as it has pure non-dispersive nature. This is precisely the origin for positive sign of the vacuum energy:  $\Delta E_{\rm vac}>0$.

\section{Conclusion}\label{conclusion}
The formal result of the present work can be expressed  by eqs.(\ref{ratio}), (\ref{vacuum_energy}), and we shall not repeat    the  comments listed in   the last section \ref{energy} explaining some important consequences of this result.  If  the same effect persists in FLRW Universe (\ref{FLRW}), which we expect to be the case,  it may have a number of profound consequences for understanding of the past, present  and future evolution of our Universe. 

First of all, 
the nontrivial holonomy (\ref{holonomy})   implies   the presence of $ \mathbb{S}^1$ as a part of our   space-time of our Universe. It is an additional  invariant characteristic of the manifold which can not be reduced to the local curvature. In construction discussed in the present work the corresponding   $ \mathbb{S}^1$ is identified with Euclidean time direction. This is because    the original intention in early works  on the subject (where the corresponding topological vacuum configurations were invented) was to analyze  the temperature dependence of the QCD phase transition. In principle, similar $ \mathbb{S}^1$ could be also a part of spatial coordinates. 
Such an assumption definitely consistent with all known  observations if the size of  the $ \mathbb{S}^1$ is sufficiently large $\gtrsim H^{-1}$   at present epoch, see \cite{Urban:2009ke} for the estimates in the given context. Furthermore, the linear correction enters formula (\ref{vacuum_energy}) in form of its absolute value $|\kappa|$   as it essentially describes the linear  (positively defined) size of the corresponding manifold. In context of  FLRW Universe a similar statement implies that the linear correction in $H$ enters  formula (\ref{FLRW}) in form of its absolute value $|H|$, see footnote \ref{H} for terminological  clarification. Therefore, it can not lead to any T-violating effects, which  
one could suspect as $H=\dot{a}/a$  indeed is a T-odd parameter. Still, it generates the de Sitter behaviour (\ref{a}) as a result of this linear scaling. 

Another profound consequence of this framework is as follows. The conventional  scenarios of the eternal self-producing inflationary universes are always formulated in terms of a physical scalar dynamical inflaton field  $\Phi(x)$. This problem with self-reproduction of the universe does not even emerge in our framework as there are no any fundamental scalar  dynamical  fields in the system responsible for inflation.
Instead, the   de Sitter behaviour (\ref{a}) in  our framework  is pure quantum phenomenon, which     is a consequence of  the dynamics of the long ranged topological configurations with nontrivial holonomy, rather than a result of a physical fluctuating dynamical field. A ``strange nature" of this type of energy  manifests itself in terms of the ``wrong" sign in the correlation function which can not be formulated in terms of any local propagating degrees of freedom as explained in section \ref{contact}. The corresponding  topological configurations which are responsible for this behaviour   may  generate, as argued in this work,  the linear in $H$ correction in the Friedman equation (\ref{friedman-infl}) which eventually leads to the de Sitter behaviour. Other problems formulated in terms of scalar   inflaton field  $\Phi(x)$ (such as large initial value $\Phi_{\rm in}\gg M_{\rm PL}$ for the inflaton) do not emerge in this framework, see \cite{Zhitnitsky:2013pna,Zhitnitsky:2014aja} for the details. 

Finally, we should   mention that  the energy described by a formula similar to eq. (\ref{FLRW}) (which  eventually leads to the de Sitter behaviour (\ref{a})) has been previously postulated \cite{dyn, 4d} as the driving force for the dark energy. The model has been (successfully) confronted with observations, see recent review papers \cite{Cai:2014pek,Cai:2012fq}   and many original references therein, where it has been claimed that this proposal is consistent with all presently available data,  see also ref. \cite{Gomez-Valent:2014fda} for completeness. Our comment here is that history of evolution of the universe may repeat itself by realizing the de Sitter behaviour twice in its history.  The $\qcd$-dynamics was responsible for the inflation in early universe, while   the QCD dynamics is responsible  for the dark energy in present epoch.

 We conclude this work (mainly devoted to analysis of the topological configurations with typical energy scale $\Lbar$) with the following comment related to  a fundamentally  different problem with drastically  different energy scale.  
 Namely, as we discussed at length in this paper, the  heart of the proposal is a fundamentally new type of energy  (\ref{FLRW}), (\ref{Z3}),  (\ref{vacuum_energy}) which can  not  be expressed in terms of  any propagating degrees of freedom. 
Rather, this novel  contribution to the energy has non-dispersive nature. The effect is formulated in terms of the tunnelling processes between topologically different but physically identical states.
This novel type of energy, in fact, has been well studied in the QCD lattice simulations in the flat background, see  \cite{Zhitnitsky:2013pna} for references on the original lattice results. 
Our comment relevant for the present study is that this fundamentally new type of energy can be, in principle, studied in a tabletop experiment by measuring some specific corrections to the Casimir vacuum energy  in the Maxwell theory as suggested in  \cite{Cao:2013na,Zhitnitsky:2013hba,Zhitnitsky:2014dra,Zhitnitsky:2015fpa}.
This fundamentally new contribution to the Casimir pressure emerges as a result of tunnelling processes, rather than due to the conventional fluctuations of the propagating photons with two physical transverse polarizations.  This effect does not occur for the scalar field theory, in contrast with conventional Casimir effect which is operational for both: scalar as well as for Maxwell fields.
The extra energy computed in \cite{Cao:2013na,Zhitnitsky:2013hba,Zhitnitsky:2014dra,Zhitnitsky:2015fpa} is the direct analog of the non-dispersive contribution to the energy   (\ref{FLRW})   which is the key player of the present work.
In fact, an extra contribution to the Casimir pressure emerges in this system as a result of nontrivial holonomy similar to (\ref{holonomy}) for the Maxwell field. The nontrivial holonomy is  enforced by  the nontrivial boundary conditions imposed
 in refs \cite{Cao:2013na,Zhitnitsky:2013hba,Zhitnitsky:2014dra,Zhitnitsky:2015fpa}.  

 \section*{Acknowledgements} 
  I am tankful to the participants of the workshop ``Future prospects for fundamental particle physics and cosmology" (Simons Center for Geometry and Physics, Stony Brook, May 2015) where this work was presented,  for their comments, questions and suggestions. 
    This research was supported in part by the Natural Sciences and Engineering Research Council of Canada.

\appendix
\section{\label{terminology}Few comments on  fractionally charged constituents and the terminology}
The constituents of the KvBLL   configurations  were originally \cite{Kraan:1998pm,Lee:1998bb} called the BPS (Bogomolnyi-Prasad-Sommerfeld)  {monopoles} (\ref{M_bc}) and KK (Kaluza- Klein)-monopoles (\ref{L_bc}) correspondingly.  
These  configurations were  later on coined as  $M$- { dyons} and $L$- dyons   to emphasize that they  carry the electric charges along with the magnetic charges \cite{Diakonov:2004jn,Diakonov:2007nv}.  This (incorrect) interpretation based on observation that  these configurations carry the topological charges (and naively the electric charges) along with magnetic charges. One should remember, however, that  the monopoles in this construction are pseudoparticles living in 4d Euclidean space-time, rather than static 3d objects. The finite action   and  finite topological charge   for these objects   results from wrapping of the monopole's path along the Euclidean time direction $ {\mathbb{S}}^1$ with nontrivial holonomy (\ref{polyakov}). 
These objects do not carry a conventional static electric charge;  nevertheless, they do carry the topological charges defined in 4d Euclidean space-time. Furthermore, the second types of the monopoles, the $L$ monopoles are time dependent configurations, and do not exist as static objects in 3d Euclidean space. Therefore, we keep notations for letters $M$ and $L$ suggested in  \cite{Diakonov:2004jn,Diakonov:2007nv}, but we use term ``monopoles" rather than ``dyons" in the present work.
When the holonomy assumes its ``confining" value $\nu=1/2$ the topological charges of the constituents assume $Q=\pm1/2$.

In more generic case of the $SU(N)$ gauge group the topological charge $Q=\pm1/N$ for ``confining" holonomy such that a single KvBLL   configuration
can be thought as a superposition of $N$ different types of monopoles which carry $N$ different types of magnetic charges and fractional $1/N$ topological charge such that the superposition carries an integer topological charge. 

We believe a short historical detour on fractionalization of the topological charge in QFT is warranted here. In the given context   fractional topological objects appear in 2 dimensional $CP^{N-1}$ model~\cite{Fateev} which were coined as {\it instanton quarks} (other names:{ instanton partons, fractional instantons}). These quantum objects carry fractional topological charge $Q=\pm 1/N$, and they are very similar to the  $L$ and $M$  monopoles  discussed in this work. These objects do not appear individually in path integral; instead, they appear as configurations consisting $N$ different   objects with fractional charge $1/N$ such that the total topological charge of each configuration   is always  integer. In this case $4Nk$ zero modes for $k$ instanton solution is interpreted as $4$   translation zero modes modes accompanied by  every single instanton quark. The same counting holds, in fact,  for any gauge group $G$, not limited to $SU(N)$ case. 
While the instanton quarks emerge in the path integral coherently, 
  these objects are highly delocalized: they may emerge on opposite sides of the space time or be close to each other with alike   probabilities. 
Similar objects have been discussed in a number of papers in  different contexts, including the topic of the present work,  \cite{Kraan:1998pm,Lee:1998bb,Diakonov:2004jn,Diakonov:2007nv,Yaffe:2008,Thomas:2011ee,Davies:1999uw,Zhitnitsky:2006sr,Parnachev:2008fy,Collie:2009iz,Bolognesi:2011nh,Zhitnitsky:2013wfa}.
 
  In particular, it has been argued that the well-established $\theta/N$ dependence in strongly coupled QCD (expressed by formula (\ref{theta}) for  specific case  $N=2$) unambiguously implies that the relevant configurations in QCD must carry   fractional topological charges in confinement phase, see review preprint \cite{Zhitnitsky:2006sr} and the references on earlier original results therein.      The weakly coupled deformed QCD model \cite {Yaffe:2008,Thomas:2011ee,Zhitnitsky:2013hs} where computations are under complete theoretical control  is a precise dynamical realization of this idea when the fractionally charged monopoles are responsible for confinement, saturate the  topological susceptibility with a ``wrong sign", generate the ``secrete long range forces", suspected long ago \cite{Luscher:1978rn},   and provide other crucial elements which  are known to exist  in strongly coupled regime as reviewed in section \ref{contact}.
  
  Furthermore, it has been argued in \cite{Parnachev:2008fy,Zhitnitsky:2013wfa} that the confinement deconfinement phase transition within this framework can be interpreted as Berezinskii-Kosterlitz-Thouless (BKT) -like phase transition:  at $T> T_{c}$
   the   constituents prefer to organize a single  caloron  of a finite size.  We coin this phase as a  ``molecular phase" which corresponds to a de-confined phase
   in conventional terminology. When one crosses the  phase transition line  at $T< T_{c}$
     the constituents (which are called $L, M$ monopoles in the present work) prefer to stay far away 
     from each other.
   It corresponds to the dissociation of each caloron into $N$ constituents, and we 
   call this state  as the   ``$N$ component plasma phase" in 4d Euclidean space. This regime corresponds to  the confined phase
   in conventional terminology when all constituents  are delocalized in 4d Euclidean space. The gap in this confined phase is determined by the Debye correlation length of this 4d   plasma. The arguments  \cite{Parnachev:2008fy,Zhitnitsky:2013wfa}  are based on large $N$ counting, but we believe that this picture holds  for any finite $N$. Recent numerical studies \cite{Liu:2015ufa,Liu:2015jsa,Larsen:2015vaa} are capable, in principle,  to bring these  large $N$ qualitative arguments  into  a solid theoretical framework.

\end{document}